\documentclass[10pt,twocolumn]{article}

\usepackage{graphics}
\usepackage{graphicx}
\usepackage{enumerate}
\usepackage{stfloats}
\usepackage[top=0.5in, bottom=1in, left=0.75in, right=0.75in]{geometry}
\usepackage{natbib}
\usepackage{titlesec}

\titleformat{\section}
  {\normalfont\fontsize{12}{15}\bfseries}{\thesection}{0.5em}{}
\titleformat{\subsection}
  {\normalfont\fontsize{10}{13}\bfseries}{\thesubsection}{0.5em}{}
\begin{document}
\twocolumn[
{\centering{} 

\textbf{\large Electromagnetic stabilization of tokamak microturbulence in a high-$\beta$ regime}
\vspace{1mm}

J.~Citrin$^{1,2}$, J. Garcia$^2$, T. G\"{o}rler$^3$, F.~Jenko$^3$, P.~Mantica$^4$, D.~Told$^3$, C.~Bourdelle$^2$, D.R. Hatch$^5$, G.M.D.~Hogeweij$^1$, T.~Johnson$^6$,  M.J.~Pueschel$^7$, M. Schneider$^2$, and JET-EFDA contributors$^*$

\vspace{1mm}
JET-EFDA, Culham Science Centre, Abingdon, OX14 3DB, UK
\footnotesize
\vspace{1mm}
\textit{ \\$^1$FOM Institute DIFFER -- Dutch Institute for Fundamental Energy Research, PO Box 1207, 3430 BE Nieuwegein, The Netherlands\\ $^2$CEA, IRFM, F-13108 Saint Paul Lez Durance, France \\ $^3$Max Planck Institute for Plasma Physics, Boltzmannstr. 2, 85748 Garching, Germany \\ $^4$Istituto di Fisica del Plasma ``P. Caldirola'', Milano, Italy \\
$^5$Institute for Fusion Studies, University of Texas at Austin, Austin, Texas 78712, USA
$^6$EES, KTH, Stockholm, Sweden \\ $^7$Department of Physics, University of Wisconsin-Madison, Madison, Wisconsin 53706, USA \\ $^*$See the Appendix of F. Romanelli et al., Proceedings of the 24th IAEA Fusion Energy Conference 2012, San Diego, USA} }
\vspace{2mm}

\small \begin{changemargin}{1cm}{1cm} The impact of electromagnetic stabilization and flow shear stabilization on ITG turbulence is investigated. Analysis of a low-$\beta$ JET L-mode discharge illustrates the relation between ITG stabilization, and proximity to the electromagnetic instability threshold. This threshold is reduced by suprathermal pressure gradients, highlighting the effectiveness of fast ions in ITG stabilization. Extensive linear and nonlinear gyrokinetic simulations are then carried out for the high-$\beta$ JET hybrid discharge 75225, at two separate locations at inner and outer radii. It is found that at the inner radius, nonlinear electromagnetic stabilization is dominant, and is critical for achieving simulated heat fluxes in agreement with the experiment. The enhancement of this effect by suprathermal pressure also remains significant. It is also found that flow shear stabilization is not effective at the inner radii. However, at outer radii the situation is reversed. Electromagnetic stabilization is negligible while the flow shear stabilization is significant. These results constitute the high-$\beta$ generalization of comparable observations found at low-$\beta$ at JET. This is encouraging for the extrapolation of electromagnetic ITG stabilization to future devices. An estimation of the impact of this effect on the ITER hybrid scenario leads to a 20\% fusion power improvement. \end{changemargin}\vspace{6mm} ] 

\section{Introduction}
\label{sec:intro}

The ion-temperature-gradient (ITG) instability~\cite{roma89,guo93} is well established as a primary driver for ion and electron turbulent heat transport in tokamak plasmas~\cite{ITER2}. While long studied, open questions remain on aspects of ITG turbulence. Specifically, experimental observations in JET tokamak L-mode discharges of significantly reduced ion temperature profile stiffness~\cite{mant09,mant11} within the tokamak plasma half-radius -- hitherto unexplained by direct numerical simulation -- have recently triggered intense investigation. ``Stiffness'' is defined here as the gradient of the gyroBohm normalized ion heat flux with respect to the driving normalized logarithmic ion temperature gradient $R/L_{T_i}$. While the original experimental hypothesis posed a connection between the reduced stiffness and concomitant low magnetic shear and high $E{\times}B$ flow shear, this was not borne out by dedicated nonlinear gyrokinetic simulations. Rather, following an extensive numerical study using the \textsc{Gene} gyrokinetic code~\cite{jenk00}, the physical mechanism able to explain the observations is nonlinear electromagnetic stabilization of ITG turbulence~\cite{citr13,citr14}. 

\textit{Linear} stabilization of ITG modes by electromagnetic (EM) effects (i.e., finite-$\beta$) is well known~\cite{weil92,hiro00}. However, recent \textit{nonlinear} simulations of finite-$\beta$ ITG turbulence have highlighted that nonlinear EM-stabilization -- i.e. the relative reduction of the heat fluxes compared with the electrostatic (ES) case -- is significantly greater than the relative reduction of the linear growth rates between the EM and ES cases~\cite{pues08,pues10}. This is correlated with an increase in zonal flow drive in the EM case~\cite{pues13b}. 

When including fast ions in the system, the suprathermal pressure gradients further increase the EM coupling, augmenting the stabilization effect. This was first observed in linear calculations~\cite{roma10}. The nonlinear enhancement of the total EM-stabilization over the linear stabilization is maintained in the system including fast ions. In Refs.~\cite{citr13,citr14} this was key to explain the low $T_i$-stiffness observations in Refs.\cite{mant09,mant11}. There, the low stiffness plasmas were dominated by suprathermal pressure within half-radius. This finding has positive ramifications for extrapolation to ITER and reactor performance, where rotation is expected to be low but fusion $\alpha$-particles will provide a significant source of suprathermal pressure. 

Since the discharges studied in Refs.~\cite{citr13,citr14} were relatively low-$\beta$ ($\approx 0.5\%$), a natural question arises: does this physics change in more reactor-relevant, high-$\beta$ plasmas? An investigation of this question is the topic of the present paper.  A JET hybrid scenario, 75225 (with C-wall)~\cite{hobi12}, was chosen for detailed linear and nonlinear gyrokinetic study. The importance of fast ions on ITG stabilization in this discharge is already seen in previous linear studies~\cite{garc13}. In general, the enhancement of plasma diamagnetism by the suprathermal pressure gradient is believed to be essential for attaining hybrid regimes~\cite{garc10}. This encouraged the comparison of the ITG stabilization physics in this discharge with the aforementioned L-mode studies. Hybrid scenarios are natural candidates for such analysis due to: a significant suprathermal pressure fraction, arising from relatively low plasma density; low magnetic shear in the inner half-radius, experimentally observed to be correlated with reduced $T_i$ stiffness~\cite{mant11b}; and no deleterious MHD activity, which otherwise would interfere with microturbulence transport studies. The nonlinear studies shown here constitute an extension of results shown in Ref.~\cite{garc14}, which also discusses the importance of fast ions for edge confinement improvement, leading to a beneficial core-edge coupling. The importance of fast ion enhanced linear EM-stabilization in this discharge is also predicted by the TGLF quasilinear transport model~\cite{stae05} when applied in an integrated modeling framework~\cite{baio14}.

This paper concentrates on the comparison between well resolved and extensive direct numerical simulation with experimental power balance fluxes, and the ramifications thereof on relevant transport questions and extrapolations to future devices. The fundamental theory of the physics mechanisms discussed here is to a large extent still an open question, and left for future work. 

The rest of the paper is organized as follows. In section~\ref{sec:settings} the simulation and discharge parameters are reviewed. Section~\ref{sec:param} discusses the relevant parametrization of the EM-stabilization. Section~\ref{sec:linear} reviews the linear analysis of discharge 75225. Section~\ref{sec:nonlinear} describes the nonlinear simulations of discharge 75225. Conclusions are presented in section~\ref{sec:discuss}.

\section{Simulation setup and discharge parameters}
\label{sec:settings}

\begin{table*}[tp]
\small
\centering
\caption{\footnotesize Discharge 75225 dimensionless parameters as input into the $\textsc{Gene}$ simulations. The values are averaged between 6.0-6.5 s, before the onset of a deleterious $n=3$ neoclassical tearing mode. $\nu^*$ is the normalized collisionality: $\nu^*{\equiv}\nu_{ei}\frac{qR}{\epsilon^{1.5}v_{te}}$, with $\epsilon=a/R$ and $v_\mathrm{te}\equiv\sqrt{T_e/m_e}$. $\gamma_E$ is the $E{\times}B$ shear rate, where $\gamma_E{\equiv}\frac{r}{q}\frac{d\Omega}{dr}/(\frac{v_\mathrm{th}}{R})$ for the purely toroidal rotation assumed here for the NBI driven cases. $v_\mathrm{th}\equiv\sqrt{T_e/m_i}$. The gradient lengths are defined taking the radial coordinate as the toroidal flux coordinate. We note that the $\beta_j$ of all ionic species are scaled with $\beta_e$, according to the pressure ratios $p_j/p_e$.}
\tabcolsep=0.11cm
\scalebox{0.9}{\begin{tabular}{c|c|c|c|c|c|c|c|c|c|c|c|c|c|c}
\label{tab:summary}
Location & $R/L_{Ti}$ & $R/L_{Te}$ & $R/L_{ne}$ & $T_i/T_e$ & $\beta_e$ [\%] & $\hat{s}$ & $q$ & $\nu^*$ & $Z_{\mathrm{eff}}$ & $\gamma_E$ & $n_{\mathrm{fast}}/n_e$  & $T_{\mathrm{fast}}/T_e$ & $R/L_{T\mathrm{fast}}$ & $R/L_{n\mathrm{fast}}$ \\
\hline
$\rho=0.33$ & 6.2 & 3.4 & 2.4  & 1.2  & 1.8  & 0.16 & 1.1  & 0.023 & 1.55 & 0.2 &  0.12  & 7.3 & 0.33 & 7.0 \\
$\rho=0.64$ & 3.4 & 5.8 & 3.05 & 1.07 & 0.75 & 1.44 & 1.74 & 0.018 & 1.90 & 0.3 &  0.055 & 8.65& 2.15 & 8.9 \\
\hline
\end{tabular}}
\end{table*}

\textsc{Gene} solves the gyrokinetic Vlasov equation, coupled self-consistently to Maxwell's equations, within a ${\delta}f$ formulation. $\textsc{Gene}$ works in field line coordinates, where $x$ is the radial coordinate, $z$ is the coordinate along the field line, and $y$ is the binormal coordinate. Both an analytical circular geometry model (derived in Ref.\cite{lapi09}) as well as a numerical geometry were used in this work. All simulations carried out were local. Collisions are modeled using a linearized Landau-Boltzmann operator. Both linear and nonlinear initial value simulations were performed. Typical grid parameters were as follows: perpendicular box sizes $[L_x,L_y]=[250,125]$ in units of ion Larmor radii, perpendicular grid discretisations $[n_x,n_\mathrm{ky}]=[256,32]$, 32 point discretisation in the parallel direction, 48 points in the parallel velocity direction, and 12 magnetic moments. The high $L_x$ and $n_x$ values were necessary to satisfy the boundary conditions in the low magnetic shear simulations at low radii, and were relaxed for simulations with higher magnetic shear. Parallel magnetic fluctuations were included in the simulations of JET discharge 75225. This cannot be neglected, due to the relatively high $\beta$, and is important for setting the strength of electromagnetic coupling, as seen in dedicated checks.

JET discharge 75225 was analyzed at two separate radial locations, at $\rho=0.33$ and $\rho=0.64$, where $\rho$ is the normalized toroidal flux coordinate. The discharge is characterized by a peaked $T_i$ profile in the inner half-radius, motivating this choice of separating the analysis to the two separate regions, and examining the relative impact of EM-stabilization and $E{\times}B$ flow shear stabilization in each region. For preparing the $\textsc{Gene}$ input, interpretative simulations of the discharge with the CRONOS integrated modeling suite~\cite{arta10} were carried out. These simulations included calculations of the NBI heat deposition and fast ion profile using the NEMO/SPOT code~\cite{schn11}, power balance analysis, current diffusion, and magnetic equilibrium calculations with HELENA~\cite{huys91}. The dimensionless parameters for the discharge in the two regions, included as input for the $\textsc{Gene}$ simulations, are shown in Table.~\ref{tab:summary}. The fast ions were approximated in $\textsc{Gene}$ as hot isotropic Maxwellians, taking the average $T_\mathrm{fast}$ from the calculated slowing down distribution.

\section{Parametrization of EM-stabilization}
\label{sec:param}

\begin{figure*}[htbp]
	\centering
		\includegraphics[scale=0.52]{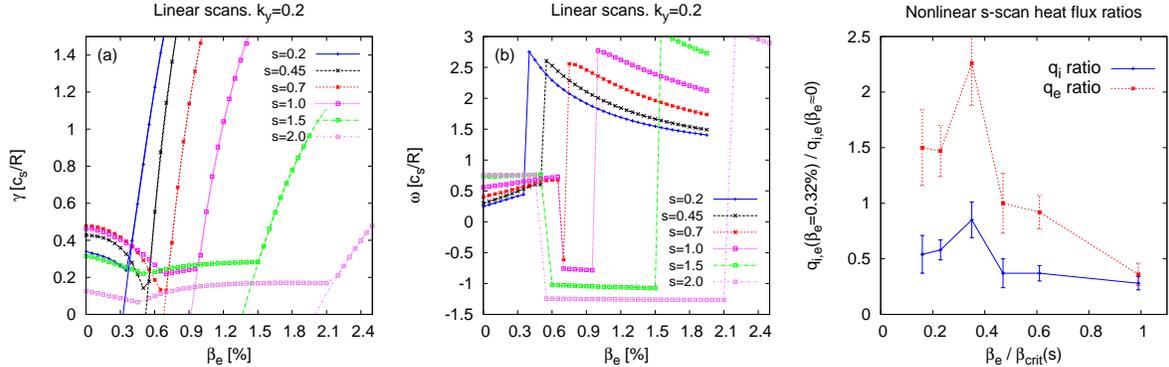}
		\caption{\footnotesize $\beta$ scans for various $\hat{s}$ values for the low-stiffness JET discharge 66404, at $\rho=0.33$. The nominal $\beta_e$ and $\hat{s}$ is 0.32\% and 0.7 respectively. The left panel (a) shows the linear growth rates and the quadratic fit to the $\beta_\mathrm{crit}$ KBM threshold (the curves meeting the x-axis). The center panel (b) shows the corresponding frequencies, and the right panel (c) shows the ratio between the EM-case and ES-case heat fluxes from the corresponding nonlinear simulations.}
	\label{fig:sscan}
\end{figure*}

Before proceeding to the analysis of JET discharge 75225, we first discuss the parametrization of the strength of the linear and nonlinear EM-stabilization. This is to gain intuition on the relevant parameter regimes of the effect. Since the linear stabilization of the toroidal ITG mode is closely linked with the increase of the electromagnetic instability drive~\cite{kim93}, we take $\beta/\beta_\mathrm{crit}$ as a valid parameter of merit. This will be justified in this section. 

$\beta_\mathrm{crit}$ is the kinetic $\beta$ limit of the EM ion mode, e.g., Kinetic Ballooning Modes (KBM), fast ion driven Alfv\'{e}nic modes, or more generally 
a hybrid case. $\beta_\mathrm{crit}$ is a function of the plasma parameters, including the driving gradient lengths. In the fluid limit the stability threshold is parametrized by $\alpha_\mathrm{MHD}$, where $\alpha_\mathrm{MHD}{\equiv}q^2R{\beta}'=q^2\sum_j\beta_j\left(\frac{R}{L_{Tj}}+\frac{R}{L_{nj}}\right)$, summing over all species. $\beta_\mathrm{crit}$ and $\alpha_{crit}$ are thus intrinsically linked, where $\alpha_{crit}$ effectively captures a subset of the parameter dependencies in $\beta_\mathrm{crit}$. These parametrizations approximately carry over into the kinetic system~\cite{pues08}, and thus an increase of the ITG stabilization with pressure gradients is expected (through $\beta'$). This is the key factor underlying the mechanism of fast ion stabilization of ITG, since the suprathermal pressure can increase the total pressure gradient (and thus $\beta'$) and increase the stabilization factor, while not simultaneously adding to the underlying ITG drive. We stress that this effect is independent from the impact of the pressure profile on the magnetic drift frequency ($\omega_d$), and from self-consistent modifications of the magnetic equilibrium. The former can be particularly effective at low magnetic shear~\cite{bour05} and in $\hat{s}-\alpha$ geometry is directly linked to the $\alpha_\mathrm{MHD}$ dependence of $\omega_d$. In all cases discussed in this section, the analytical unshifted circular geometry formulation was used, with thus no impact of $\beta'$ on the magnetic equilibrium. While the ${\nabla}P$ in the $\omega_d$ calculation did scale with $\beta$, test cases with a constant ${\nabla}P=0$ in the $\omega_d$ calculation were run. Only small differences ($<$10\%) in $\beta_\mathrm{crit}$ were observed, with the same qualitative behavior throughout. 

A study of a subset of the $\beta_\mathrm{crit}$ dependencies was carried out for the low-stiffness low-$\beta$ JET discharge 66404 at $\rho=0.33$, discussed in detail in Ref.\cite{citr13,citr14,ryte11}. Of particular interest is the magnetic shear ($\hat{s}$) dependence, motivated by the observation of increased destiffening at low-$\hat{s}$~\cite{mant11b}.  In figure~\ref{fig:sscan} a $\beta$ scan is shown for various levels of $\hat{s}$, where $\hat{s}=0.7$ is the nominal value for the discharge. These scans include both linear calculations and corresponding nonlinear simulations. In all cases here, only the thermal species are included, although nominally 66404 has a significant fast ion population. This is an extension of the $\hat{s}$-scan for this discharge shown in Refs.\cite{citr13,citr14}. The linear runs where carried out at $k_y=0.2$, the wavenumber corresponding to the most unstable KBM mode. $k_y$ is normalized by $1/\rho_s$

For all values of $\hat{s}$, a transition from ITG modes to KBM is observed, characterized by a sharp upswing of the growth rates for increasing $\beta$ and a jump in the real frequencies of the dominant mode. Positive $\omega$ is defined as the ion diamagnetic direction in $\textsc{Gene}$. For $\hat{s}>1$, an intermediate TEM zone is apparent, characterized by the negative frequencies. The KBM thresholds for each case were extrapolated by a quadratic fit to the first several KBM growth rates, shown in figure~\ref{fig:sscan}a. 

An $\hat{s}$-scan of nonlinear simulations -- both EM and ES -- was then carried out, with the nominal $\beta_e=0.32\%$ corresponding to discharge 66404 at $\rho=0.33$. The ratio between the EM-case and ES-case heat fluxes at each point in the scan is shown in figure~\ref{fig:sscan}c, plotted against the $\beta_e/\beta_\mathrm{crit}$ for each parameter set, where $\beta_\mathrm{crit}$ was taken from the fit of the linear scans. Two important points should be emphasized: a) the $\beta$ scans and $\beta_\mathrm{crit}$ shown here correspond to electron $\beta$, but the $\beta_s$ of all ionic species in the simulations are scaled self-consistently with $\beta_e$, according to the pressure ratios $p_j/p_e$; b) the $\hat{s}$-dependent $\beta_\mathrm{crit}$ is the parameter which changes within the scan in figure~\ref{fig:sscan}c, not $\beta_e$. While the intermittency in the simulations leads to significant error bars, a clear separation in this flux ratio is seen between high and low $\beta_e/\beta_\mathrm{crit}$, with a lower ratio at higher $\beta_e/\beta_\mathrm{crit}$. This supports the adoption of $\beta/\beta_\mathrm{crit}$ as a valid parameter of merit also for the nonlinear ITG EM-stabilization, suggesting that the EM modification of the linear modes is also linked to the enhanced nonlinear stabilization. For the ion heat flux, the inclusion of EM effects is always stabilizing, consistent with ITG stabilization. However, the situation is more complicated with the electron heat flux. At higher $\hat{s}$ (higher $\beta_\mathrm{crit}$ and thus lower $\beta_e/\beta_\mathrm{crit}$), the inclusion of EM effects leads to an increase of electron heat flux in this specific case. This is due to a combination of magnetic flutter transport, and likely an increased relative impact of the trapped electron drive. More detailed analysis is necessary to untangle the various effects. We note however that all these cases remain ion heat flux dominated, and even for the EM simulations, the $q_i/q_e$ ratio remains within the range of 2-4.

\begin{figure}[htbp]
	\centering
		\includegraphics[scale=0.42]{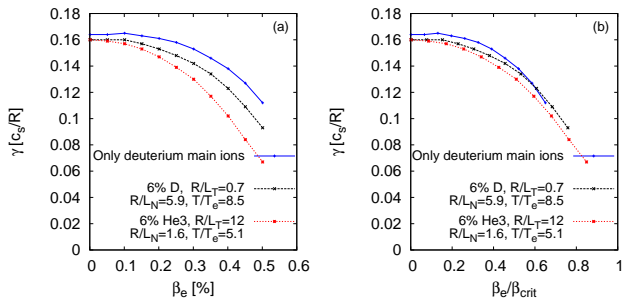}
		\caption{\footnotesize Linear $\beta$ scans for JET discharge 66404 at $\rho=0.33$, for thermal species only and when including either NBI or ICRH driven fast ions. The scans are plotted as a function of $\beta_e$ (a) and as a function of $\beta_e/\beta_\mathrm{crit}$ (b), where $\beta_\mathrm{crit}$ was separately calculated for each case.}
	\label{fig:bscan}
\end{figure}

In figure~\ref{fig:bscan}, the relevance of the $\beta/\beta_\mathrm{crit}$ parametrization of EM-stabilization in the presence of fast ions is examined. Figure~\ref{fig:bscan}a displays a $\beta$-scan for three separate cases of discharge 66404 at $\rho=0.33$: with thermal species only, with a fast D species matching the modeled NBI profile for the discharge, and with a fast $^3\mathrm{He}$ species matching the modeled ICRH profile. In this scan, a separation of the curves are evident, with the EM-stabilization occurring at different rates with increasing $\beta_e$. Continuing the calculation of the curves down to the KBM limit (not shown for brevity) we obtain $\beta_\mathrm{crit}=0.77\% ,0.66\%, 0.59\%$ respectively for the three cases. When rescaling the x-axis of the plot to $\beta_e/\beta_\mathrm{crit}$, as shown in figure~\ref{fig:bscan}b, the overlap between the three curves is much more apparent, illustrating the suitability of $\beta/\beta_\mathrm{crit}$ as a parameter of merit. 

To summarize, $\beta/\beta_\mathrm{crit}$ is shown to be a valid parameter of merit for both the linear and nonlinear EM-stabilization. $\beta_\mathrm{crit}$ is reduced at lower-$\hat{s}$ and with increased suprathermal pressure gradients. Thus, for a given thermal $\beta$, enhanced EM--stabilization of ITG turbulence is expected at both low-$\hat{s}$ and high suprathermal pressure. This statement is relevant for the interpretation of transport in JET discharge 75225, as shown in the following sections. 

We conclude this section with a remark on the nature of $\beta_{crit}$. We have assumed that $\beta_{crit}=\beta_{KBM}$. However, an additional $\beta$-limit is apparent from numerical simulations -- $\beta_{NZT}$ (non-zonal-transition) -- leading to a sharp rise in flux due to a reduction in zonal flow activity~\cite{pues13,pues13c}. For highly driven systems, this occurs for $\beta_{NZT}<\beta_{KBM}$. However, for experimental parameter sets studied thus far, including all those studied in this paper, $\beta_{KBM}<\beta_{NZT}$. This is an encouraging observation since we have seen that access to $\beta\sim\beta_{KBM}$ has positive ramifications for ITG stabilization. If this observation can be shown to be general, this would mean that a prediction to the degree of EM-stabilization in future devices such as ITER could be carried out from linear analysis of the projected profiles, and determining $\beta/\beta_\mathrm{crit}$ where $\beta_\mathrm{crit}$ is taken as $\beta_{KBM}$. This is left for future work.

\section{Linear simulations of JET 75225}
\label{sec:linear}
In this section we describe the linear analysis of the JET high-$\beta$ hybrid discharge 75225. The analysis is split into 2 subsections concentrating on different radii, $\rho=0.33$ and $\rho=0.64$. 
\subsection{Linear calculations at $\rho=0.33$}
\begin{figure}[htbp]
	\centering
		\includegraphics[scale=0.55]{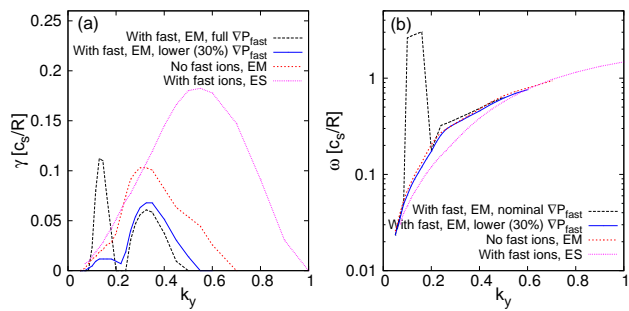}
		\caption{\footnotesize Linear growth rate (a) and frequency (b) spectra as a function of $k_y$, for JET 75225 at $\rho=0.33$. The calculations included numerical geometry, collisions, C impurities, EM fluctuations and fast ions, unless otherwise noted.}
	\label{fig:lin033}
\end{figure}

In figure~\ref{fig:lin033} we plot the $k_y$ spectrum of the linear growth rates and frequencies at $\rho=0.33$. The plot shows 4 separate cases: a case without fast ions, a case including the full fast ion pressure gradient, a case with the fast ion pressure gradient reduced by 30\%, and a case where $\beta_e$ was set to $10^{-5}$, which is effectively an electrostatic case.  The spectrum in all cases is dominated by ITG modes, characterized by the positive frequencies in the diamagnetic drift frequency range. From the comparison of the EM and ES cases, a strong linear EM-stabilization is evident, particularly at the higher $k_y$ values. In all scans shown here, the ${\nabla}P$ taken for the $\omega_d$ calculations are set to be consistent with the input $\beta$ and logarithmic gradients in each run. We also utilize two separate input geometry files, for the cases with and without fast ions respectively. These were calculated from the Grad-Shafranov solver in the CRONOS interpretative runs for each case, at the nominal $R/L_{Ti}$, where the suprathermal pressure component was removed when calculating the `no fast ion' magnetic equilibrium. 

An additional mode is clearly apparent for the parameters corresponding to including the full suprathermal pressure, at $k_y<0.2$, at significantly higher frequencies than the ITG range. This mode is stabilized by lowering the fast ion pressure gradient by 30\%, illustrating the fast ion drive of this mode. From a Rosenbluth-Hinton test~\cite{rose98} it was determined that the frequency of these modes are within 5\% of the GAM frequency for these parameters. This supports the identification of this mode as a Beta induced Alfv\'{e}n Eigenmode (BAE)~\cite{zonc96}, which is known to be degenerate with the GAM frequency~\cite{nguy08}. However, these modes are likely coupled with the KBM modes which share a similar frequency range. Thus, in lieu of deeper analysis regarding the precise characterization, in the following we refer to this mode as a hybrid BAE/KBM. As will be shown in section~\ref{sec:nonlinear}, the heat fluxes in the simulation of the system with unstable BAE/KBMs is highly inconsistent with the experimental power balance values. Thus, it is assumed that the system self-organizes to a state just below the BAE/KBM stability boundary. We maintain the lowered fast ion pressure for all the nonlinear simulations, and consider this the nominal parameter set.

\begin{figure}[htbp]
	\centering
		\includegraphics[scale=0.42]{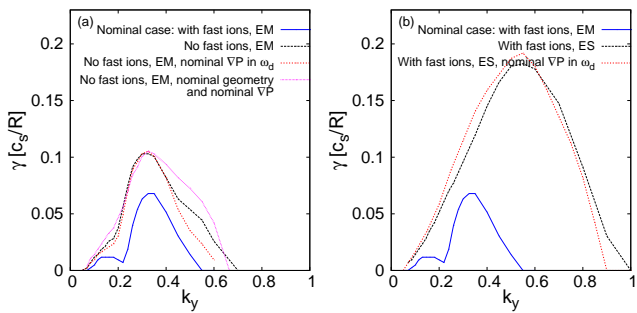}
		\caption{\footnotesize Linear growth rate spectra as a function of $k_y$, for JET 75225 at $\rho=0.33$. The impact of EM-stabilization is isolated when comparing cases with and without fast ions (a), and when comparing the EM and ES cases (b).}
	\label{fig:lin033alt}
\end{figure}

Comparing the ITG growth rates of the cases with and without fast ions shows that the fast ions provide an additional stabilization of ITG beyond the thermal EM-stabilization, as expected from section~\ref{sec:param}. This stabilization has been isolated to be electromagnetic in nature, and not due to the concomitant modifications of ${\nabla}P$ in the $\omega_d$ calculation, or the different magnetic equilibrium. This is seen in figure~\ref{fig:lin033alt}a, where the nominal `with fast ion' case is compared to various `no fast ion' cases. The ${\nabla}P$ in $\omega_d$ and then the magnetic equilibrium are respectively modified to equate with those of the nominal case. The growth rate differences are minor compared with the EM-stabilization. Interestingly, switching to the higher shifted magnetic equilibrium of the `with fast ion' case leads to destabilization, i.e. increased growth rates. While increased Shafranov shift typically stabilizes the ITG drive, in this case it also stabilizes the KBM/BAE drive and increases $\beta_\mathrm{crit}$, thus reducing the degree of EM-stabilization of ITG, leading to a net effect of higher growth rates. This impact of geometry on the KBM is linked with the observed improvement of peeling-ballooning stability in the tokamak edge region in cases with fast ions, which leads to a beneficial core-edge transport coupling~\cite{garc14}. In figure~\ref{fig:lin033alt}b we also isolate the EM-stabilization effect when comparing the EM and ES cases with fast ions. By replacing the ${\nabla}P\approx0$ in the ES case $\omega_d$ calculation (due to setting ${\nabla}P$ with respect to the input $\beta\approx0$ and gradients) with the nominal ${\nabla}P$ from the EM case, we only obtain a minor difference in calculated growth rates. The input magnetic equilibrium was already the same for both cases.

\begin{figure}[htbp]
	\centering
		\includegraphics[scale=0.56]{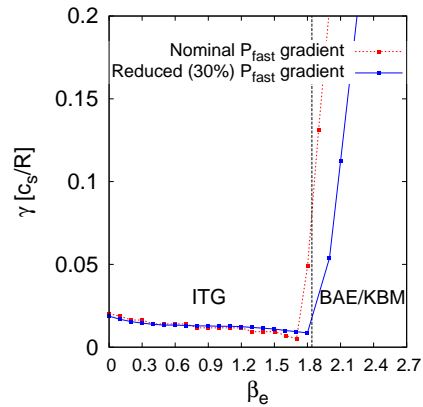}
		\caption{\footnotesize $\beta$-scan of linear growth rate at $k_y=0.1$ for JET 75225 at $\rho=0.33$. Two separate cases are shown, one with nominal parameters and the second with the fast ion pressure reduced by 30\%. The nominal $\beta_e=1.84\%$ is signified by the vertical black line.}
	\label{fig:beta033}
\end{figure}

The behavior of the BAE/KBM mode destabilization at $k_y=0.1$ is shown in figure~\ref{fig:beta033}, for the two separate cases with the full and lowered fast ion pressure gradient. For both cases the ITG growth rates are stabilized with higher $\beta_e$ until the BAE/KBM is eventually destabilized. For the full gradients, this destabilization occurs for a $\beta_e$ lower than the experimental $\beta_e$ of 1.84\%. For the lowered fast ion pressure gradient, the instability boundary is just above the experimental value. Thus, $\beta/\beta_\mathrm{crit}\approx1$ in this case, and according to the results in section~\ref{sec:param} we can thus expect significant nonlinear EM-stabilization as well. We reiterate that in the scan, the $\beta_s$ of all ionic species are also scaled in the scan with $\beta_e$, according to the pressure ratios $p_j/p_e$.

\subsection{Linear calculations at $\rho=0.64$}
\begin{figure}[htbp]
	\centering
		\includegraphics[scale=0.42]{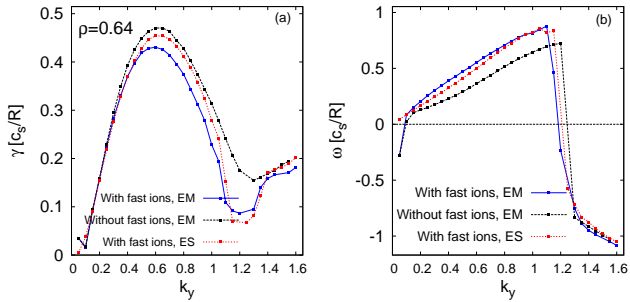}
		\caption{\footnotesize Linear growth rate (a) and frequency (b) spectra as a function of $k_y$, for JET 75225 at $\rho=0.64$. The calculations included numerical geometry, collisions, C impurities, EM fluctuations and fast ions, unless otherwise noted.}
	\label{fig:lin064}
\end{figure}

In figure~\ref{fig:lin064} we show the linear calculations at $\rho=0.64$. Three cases are shown: the nominal case including fast ions, a case without fast ions, and an electrostatic case. Three distinct instabilities are apparent. In the EM cases, a micro-tearing-mode (MTM) -- identified from its mode parity and structure -- is unstable at $k_y=0.05$. For intermediate $k_y$ ITGs are unstable, and for $k_y>1.1$ trapped electron modes (TEMs) are unstable. As opposed to the $\rho=0.33$ case, the EM-stabilization is much weaker in the lower, transport driving $k_y$ values. This is reflected by the relatively lower $\beta/\beta_\mathrm{crit}\approx0.3$ values calculated for this case.

We note that these linear calculations were carried out for modified input parameters compared with table~\ref{tab:summary}, with $R/L_{Te}$ and $R/L_{ne}$ reduced by 20\%, and $R/L_{Ti}$ increased by 20\%. This was necessary to reach agreement with the experimental power balance values and $q_i/q_e$ heat flux ratio in the nonlinear simulations. For nominal parameters, a strongly driven TEM regime was predicted at all $k_y$, leading to $q_e>q_i$, in disagreement with the observations. This modification of the driving gradients is at the limits of the experimental error bars.

\section{Nonlinear simulations of JET 75225}
\label{sec:nonlinear}
In this section we describe the nonlinear simulations of JET discharge 75225, split into 2 subsections concentrating on different radii, $\rho=0.33$ and $\rho=0.64$. 
\subsection{Nonlinear simulations at $\rho=0.33$}
\begin{figure*}[htbp]
	\centering
		\includegraphics[scale=0.75]{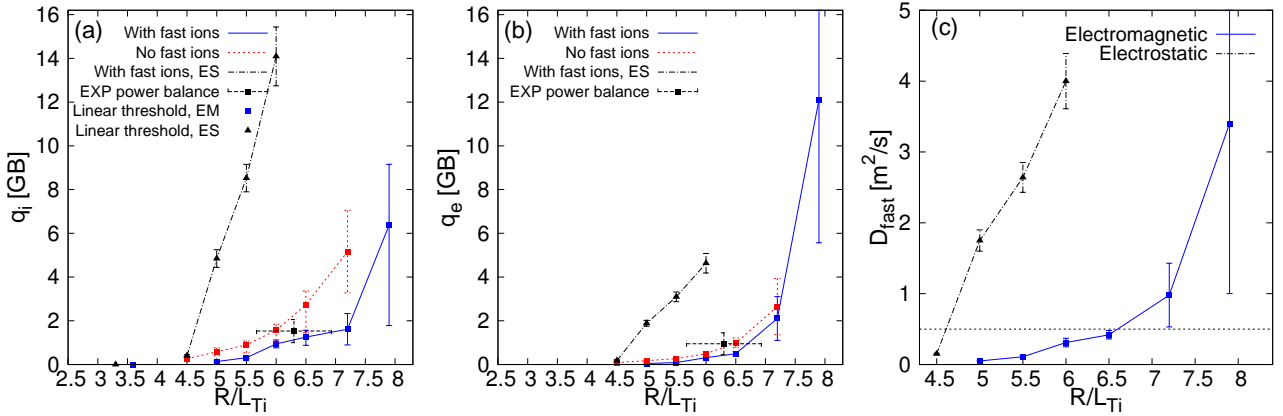}
		\caption{\footnotesize Ion heat fluxes (a), electron heat fluxes (b), and fast ion particle diffusivity (c) from \textsc{Gene} nonlinear simulations of JET 75225 at $\rho=0.33$.}
	\label{fig:nonlin033}
\end{figure*}

In figure~\ref{fig:nonlin033} the $\textsc{Gene}$ saturated nonlinear transport fluxes are shown for various cases: with fast ions (with 30\% reduced suprathermal pressure gradient), without fast ions, and with fast ions but electrostatic. In addition, the experimental ion and electron heat fluxes from power balance is shown. The $\textsc{Gene}$ error bars are derived from intermittency and uncertainties in the precise saturated potential level. The power balance ion heat flux uncertainties include statistical errors, and the propagation of $T_{i,e}$ errors on the ion-electron heat exchange term. 

The most striking observation is that the nonlinear EM-stabilization is absolutely critical for achieving agreement between the $\textsc{Gene}$ and experimental flux values. The nonlinear EM-stabilization is here manifested mostly as a stiffness reduction. The approximate factor 20 reduction in heat flux is much greater than the analogous growth rate reduction, reflecting the enhanced nonlinear nature of the EM-stabilization. This result is thus a reactor-relevant high-$\beta$ extension of the results reported in Ref.~\cite{mant11,citr13}. In addition to the stiffness reduction, a slight increase in the value of the Dimits shift~\cite{dimi00} is observed between the ES and EM case with fast ions. This is seen by comparing the linear $R/L_{Ti}$ thresholds for these cases, calculated separately and shown in the figure, with the extrapolation to zero-flux of the nonlinear stiffness curves. This Dimits shift extension is consistent with previous findings of nonlinear EM-stabilization~\cite{pues08,pues10}. In this case, as opposed to the low-$\beta$ cases in Ref.~\cite{citr13}, the majority of the nonlinear EM-stabilization is due to the thermal component. This is consistent with the lower suprathermal pressure fraction in these high-$\beta$ cases compared with the low-$\beta$ cases, on the order of 30\% here. However, the addition of fast ions does still provide significant additional stabilization, leading to an additional increase of $R/L_{Ti}$ by approximately 10--20\%, for the same heat flux level.

The level of agreement between the experimental power balance and $\textsc{Gene}$ predictions is remarkable, and provides a highly encouraging validation of nonlinear gyrokinetics in the high-$\beta$ regime. This is only possible due to the increasing capacity in high performance supercomputing capabilities. For example, the simulations shown in figure~\ref{fig:nonlin033} necessitated around 4 million CPU hours of calculation time, not including extensive convergence tests.

The rightmost datapoint in figure~\ref{fig:nonlin033}, with its sharp uptick in heat fluxes at the highest $R/L_{Ti}$, is a consequence of the destabilization of the BAE/KBM mode, which is also destabilized by the main ion gradients. This is also seen by the strong increase in fast ion diffusivity for that case. The particularly sharp uptick of the electron heat flux is mostly due to increased magnetic flutter. While the decrease in $\beta/\beta_\mathrm{crit}$ at increased $R/L_{Ti}$ due to the increase in BAE/KBM drive may help maintain the low-stiffness curve at higher $R/L_{Ti}$, this high heat flux data point is an indication of the eventual saturation of the EM-stabilization mechanism due to the crossing of the BAE/KBM boundary ($\beta/\beta_\mathrm{crit}>1$).

We recall that these simulations all used a suprathermal pressure gradient reduced by 30\%. The anomalous diffusivity prescribed in the SPOT code to provide such a suprathermal pressure reduction in the CRONOS interpretative simulations was $0.5 m^2/s$. This level is actually consistent with the $\textsc{Gene}$ predicted fast ion diffusivity for the simulation at nominal $R/L_{Ti}$, which also led to heat flux agreement with power balance. Thus, the self-consistency of this assumption is verified.

\begin{figure}[htbp]
	\centering
		\includegraphics[scale=0.42]{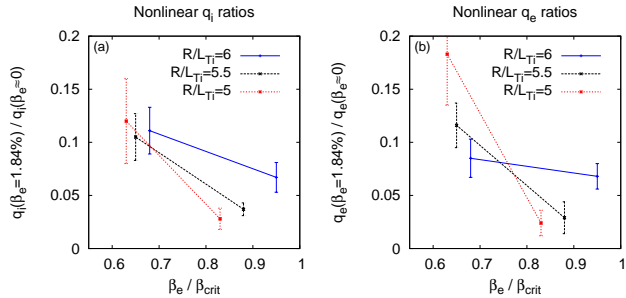}
		\caption{\footnotesize Ratio of ion heat fluxes (a) and electron heat fluxes (b) between the electromagnetic and electrostatic simulations shown in figure~\ref{fig:nonlin033}, plotted against $\beta/\beta_\mathrm{crit}$, for 3 separate $R/L_{Ti}$ inputs.}
	\label{fig:nonlinrats}
\end{figure}

The decrease in EM-stabilization when transitioning between the `with fast ions' to `no fast ions' curves in figure~\ref{fig:nonlin033}(a,b), at each given $R/L_{Ti}$ point, is correlated with a corresponding decrease in $\beta/\beta_\mathrm{crit}$. This is seen in figure~\ref{fig:nonlinrats} for both the ion and electron heat fluxes. $\beta_\mathrm{crit}$ was calculated for each case by a dedicated linear $\beta$ scan at $k_y=0.1$. This result provides further evidence of the relevance of $\beta/\beta_\mathrm{crit}$ as an organizing parameter for the impact of both linear and nonlinear EM-stabilization. However, $\beta/\beta_\mathrm{crit}$ does not correlate with $q_{i,e}(\beta_e=1.84\%)/q_{i,e}(\beta_e\approx0)$ as $\beta_\mathrm{crit}$ changes due to $R/L_{Ti}$ in the $R/L_{Ti}$ scans, particularly for the case including fast ions. This can be understood to be due to the proximity to turbulence threshold, and the complication of the increased Dimits shift of the case including fast ions. As $R/L_{Ti}$ is reduced, the case with fast ions drops to the threshold level faster than the electrostatic case, decreasing $q_{i,e}(\beta_e=1.84\%)/q_{i,e}(\beta_e\approx0)$ in spite of the lower $\beta/\beta_\mathrm{crit}$. Finally, we note that in this case, as opposed to Ref.~\cite{citr13}, the difference in flux between the `with fast ions' and `no fast ions' nonlinear EM simulations is on the same order of the difference in the linear spectra between the `with fast ions' and `no fast ions' cases in figure~\ref{fig:lin033}. This suggests that while the total degree of EM-stabilization cannot be explained purely by linear theory, the additional boost provided by fast ions can be reproduced quasilinearly \textit{for this specific case}, as confirmed in Ref.~\cite{baio14}. 

\begin{figure}[htbp]
	\centering
		\includegraphics[scale=0.32]{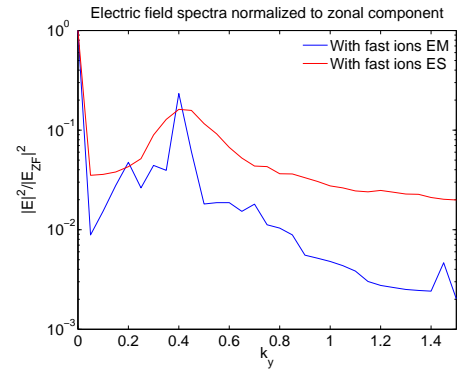}
		\caption{\footnotesize Amplitude of the normalized electric field fluctuations averaged over the saturated nonlinear phase for two separate cases (EM and ES) of JET 75225 at $\rho=0.33$. The amplitudes at each $k_y$ are averaged for all $k_x$ modes and over the parallel direction.}
	\label{fig:nonlin033_ZF} 
\end{figure}

In ITG turbulence, zonal flows (ZFs) are a well known saturation mechanism~\cite{diam05}. The correlation of the nonlinear EM-stabilization and increased destabilization of ZF secondaries has been reported~\cite{pues13b}. This connection is verified in these simulations as well. This is seen by analyzing the ratio of the electric field energy of the ZFs (all $k_y=0$ modes) to the sum of electric field energies of all drift wave modes. This value, $\left|E^2_\mathrm{ZF}\right|/\sum_{k_y>0}{\left|E^2_\mathrm{DW}\right|}$, serves a proxy for the relative impact of ZFs in the nonlinear system. This ratio is $\approx1.5$ for the EM case with fast ions, $\approx1.4$ for the EM case without fast ions, and $\approx0.65$ for the ES case. This trend correlates well with the trend of the relative heat flux ratios. In figure~\ref{fig:nonlin033_ZF}, the normalized electric field energy spectra is shown for the nonlinear simulations of the nominal EM and ES cases. The relative increased energy in the $k_y=0$ ZF modes in the EM case is clearly evident in the plot. 

\begin{figure}[htbp]
	\centering
		\includegraphics[scale=0.75]{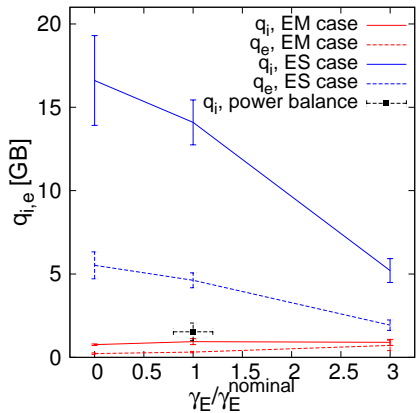}
		\caption{\footnotesize $\gamma_E$ scans for JET 75225 at $\rho=0.33$, for both EM (nominal) and assumed ES cases. The nominal $\gamma_E$ is 0.2. The simulations were carried out for $R/L_{Ti}=6$, and the ion heat fluxes are compared against the power balance value.}
	\label{fig:nonlin033_rot} 
\end{figure}

The relative role of EM-stabilization and flow shear stabilization was investigated by $\gamma_E$ scans. This is shown in figure~\ref{fig:nonlin033_rot}. The $\gamma_E$ scans, where $\gamma_E=0.2$ is the nominal value, were carried out for both EM and ES cases. A number of striking observations come out from this scan. Firstly, for the (nominal) EM case, no flow shear stabilization was evident at all. A slight increase in fluxes is even observed, particularly in the electron channel. This increased $q_e$ is due to increased magnetic flutter, which may suggest increased fast ion mode destabilization with rotation. For the ES case, a more classical trend is observed, with stabilization at increased $\gamma_E$. However, for the nominal $\gamma_E=0.2$, this stabilization remains weak. Even at $\gamma_E=0.6$, the flux levels are significantly above the power balance levels. This suggests that rotation is not an important stabilization factor in this regime, which is dominated by EM-stabilization, at least for inner radii such as $\rho=0.33$. The explanation for the complete lack of flow shear stabilization in the EM case remains unknown, and is a topic for future study. A conjecture is that this may be related to the much shorter autocorrelation times of the turbulent structures in the EM cases compared with the ES cases.

\begin{figure*}[htbp]
	\centering
		\includegraphics[scale=0.6]{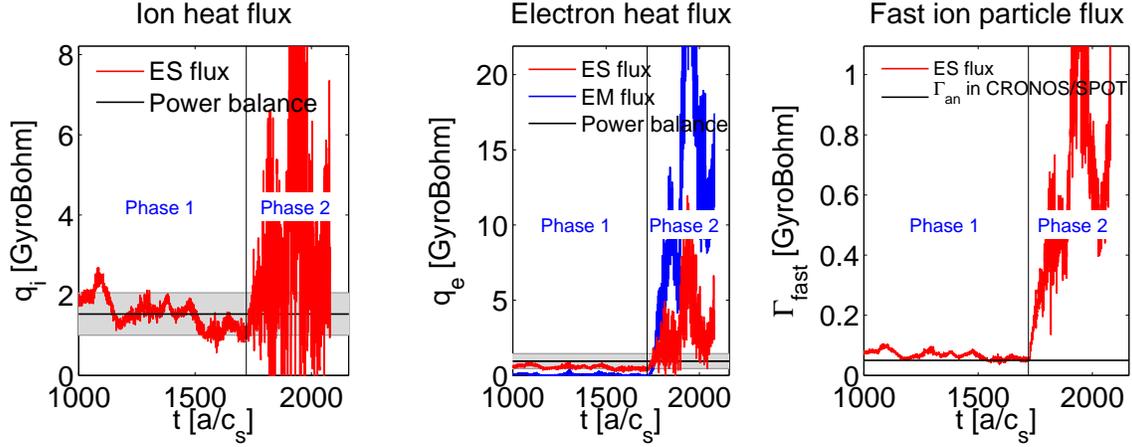}
		\caption{\footnotesize Time traces of \textsc{Gene} simulated ion heat flux (a), electron heat flux (b), and fast ion diffusivity (c) for JET 75225 at $\rho=0.33$ during a phase with a 30\% reduced fast ion pressure gradient (phase 1), followed by a transition back to the full fast ion pressure gradient (phase 2). The time traces are compared with the power balance heat fluxes, whose 1$\sigma$ confidence region is shown by the grey shaded zone.}
	\label{fig:nonlin033_BAE}
\end{figure*}

We conclude this subsection with a brief discussion of the combined ITG-BAE/KBM turbulent state. This was carried out by carrying out a simulation using the full fast pressure gradient, i.e. the case in which the BAE/KBM mode is seen to be unstable in figure~\ref{fig:lin033}. The time dependent nonlinear fluxes are shown in figure~\ref{fig:nonlin033_BAE}. In the plot, an initial phase is evident, corresponding to the case with reduced fast pressure gradient. The simulation input is then shifted to the full fast pressure gradient, and the simulation is restarted. This then corresponds to phase 2, where all fluxes rise to a degree significantly above the power balance levels. In particular, the EM component (magnetic flutter) of the electron heat flux rises sharply. This EM transport component is correlated with tearing parity fluctuations at $k_y>0.3$. MTMs are linearly stable at this region, and this observation is potentially linked to heightened nonlinear mode-mixing to tearing parity modes~\cite{hatc12} when the BAE/KBM is unstable. This is a subject of future study. However, this regime as an operating point is not experimentally relevant due to the high fluxes, and this supports the use of ``stiff'' Alfv\'{e}n Eigenmode models in reduced modeling frameworks~\cite{bass10}. Whether the system is maintained constantly below the BAE/KBM mode limit, or has limit cycle dynamics, is an open question and necessitates flux driven simulations. Thus, we cannot fully rule out the experimental relevance of this state. Finally, we note that the phase 1 time trace shown in figure~\ref{fig:nonlin033_BAE} is the tail end of a significantly longer simulation, with sufficient statistics to determine a saturated nonlinear state. However, for phase 2, there are insufficient statistics for such a determination. Nonetheless, we deem the qualitative observation of significantly higher fluxes than the power balance as robust.

\subsection{Nonlinear simulations at $\rho=0.64$}
\begin{figure}[htbp]
	\centering
		\includegraphics[scale=0.77]{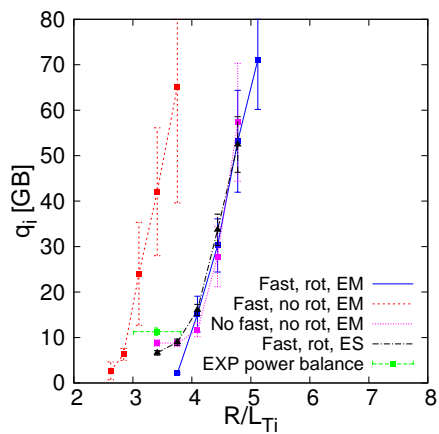}
		\caption{\footnotesize Ion heat fluxes from \textsc{Gene} nonlinear simulations of JET 75225 at $\rho=0.64$, for various cases including EM and ES simulations, and without $E{\times}B$ flow shear.}
	\label{fig:nonlin064} 
\end{figure}

In this subsection we describe the nonlinear analysis at $\rho=0.64$, and compare the trends observed with the analogous simulations at $\rho=0.33$. The $R/L_{Ti}$ scans are shown in figure~\ref{fig:nonlin064}. Numerous scans were carried out, isolating the relative roles of EM-stabilization (thermal and with fast ions), and $E{\times}B$ flow shear stabilization. It is clear that at this location, the EM-stabilization is extremely weak, with the ES and EM curves in close proximity. This is likely due to the much lower $\beta/\beta_\mathrm{crit}\equiv0.3$ in this case compared with $\rho=0.33$. 

The main observation is that flow shear stabilization is effective for these parameters, and can significantly increase the $R/L_{Ti}$ values which match the power balance fluxes. In general, the stiffness level is much higher for these parameters compared with $\rho=0.33$, likely due to the higher $\hat{s}$ and $q$ values as well as lower EM-stabilization. The $E{\times}B$ stabilization provides a threshold shift.

We reiterate that these nonlinear simulations were carried out for modified input parameters compared to the nominal parameters from table~\ref{tab:summary}, with $R/L_{Te}$ and $R/L_{ne}$ reduced by 20\%, and $R/L_{Ti}$ increased by 20\%. This was to avoid a strongly driven TEM regime and match the experimental $q_i/q_e$, and reach the vicinity of the absolute flux values.

\subsection{Extrapolation to ITER}
\begin{figure}[htbp]
	\centering
		\includegraphics[scale=0.33]{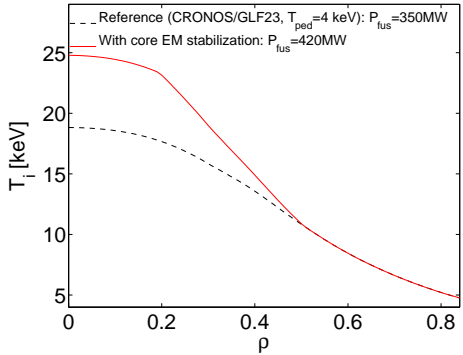}
		\caption{\footnotesize Ion temperature profiles from a reference CRONOS/GLF23 ITER hybrid scenario simulation, compared with an estimate of a peaked $T_i$ profile when including core EM-stabilization effects. The fusion powers of the two cases are compared.}
	\label{fig:ITER} 
\end{figure}

We conclude this section with a rough extrapolation of the inner core EM-stabilization effect to ITER, with the aim of judging the scale of potential impact. Theory based extrapolations carried out until now have not included the enhanced nonlinear EM-stabilization effects. This extrapolation was based on a $T_i$ profile predicted for an ITER hybrid scenario in previous theory based integrated modeling~\cite{citr10}. This modeling consisted of a CRONOS predictive simulation with the GLF23 transport model~\cite{walt97,kins05}, and a pedestal height of 4~keV. As seen in figure~\ref{fig:ITER}, this reference $T_i$ is compared with an estimate -- based on the same profile -- of $T_i$ peaking in the inner core between $\rho=0.2-0.5$. This peaking was estimated by a simple $\Delta{R/L_{Ti}}=2$ upshift, based on the results of JET 75225, which shares a similar $\beta$ and suprathermal pressure fraction as predicted for this ITER hybrid scenario reference case. The lower cutoff point of $\rho=0.2$ is based on a conservative assumption that access to the ITG EM-stabilization zone may be hindered by extremely low-$\hat{s}$ in the very inner core which decreases the $\beta_\mathrm{crit}$ of the BAE/KBMs~\cite{mora14}. Based on the $n_{D,T}$ from the reference case, the fusion power corresponding to the two separate $T_i$ profiles can be compared. The reference $P_\mathrm{fus}=350$~MW increases by 20\% to $P_\mathrm{fus}=420$~MW. The impact of core EM-stabilization is thus estimated to be significant.

\section{Conclusions}
\label{sec:discuss}

Detailed linear and nonlinear local gyrokinetic calculations of the JET high-$\beta$ hybrid discharge 75225 at $\rho=0.33$ and $\rho=0.64$ were carried out. Good agreement between the simulated and power balance heat fluxes was achieved. It was found that at $\rho=0.33$, electromagnetic (EM) stabilization of ITG turbulence is critical for reaching agreement with power balance heat fluxes, while $E{\times}B$ flow shear stabilization was not effective. Nonlinear effects and the inclusion of suprathermal pressure both significantly enhance the EM-stabilization. At $\rho=0.64$, on the other hand, the EM-stabilization was ineffective while the $E{\times}B$ flow shear stabilization was important. A likely explanation for the difference in EM-stabilization effectiveness at the two locations is the value of the $\beta/\beta_\mathrm{crit}$ EM-stabilization parameter of merit, which is $\approx$1 and 0.3 respectively for the two cases. $\beta_\mathrm{crit}$ is the critical $\beta$ value for the onset of EM instabilities. 

These results provide a high-$\beta$ extension of previous results of enhanced nonlinear EM-stabilization at low-$\beta$. The reactor relevance of this particular discharge is encouraging for extrapolations to future devices, which will have low rotation but a significant suprathermal pressure fraction due to fusion $\alpha$-particles. An estimate of the impact of this effect on ITER extrapolation was carried out, with a 20\% increase of fusion power predicted for an ITER hybrid scenario. This is a strong indication that reduced models employed for transport extrapolation to ITER and reactors must include suprathermal pressure and EM-stabilization effects validated by nonlinear gyrokinetic simulations. 

Future work will concentrate on the physical mechanism of the enhanced nonlinear EM-stabilization, which is likely related to either (or both) increased secondary ZF growth rates or reduced tertiary damping of ZFs. However, a precise characterization of the effect is still lacking.
\section{Acknowledgments}

This project has received funding from the European Union’s Horizon 2020 research and innovation programme under grant agreement number 633053. The views and opinions expressed herein do not necessarily reflect those of the European Commission. This work received financial support from NWO, and is supported by NWO-RFBR Centre-of-Excellence on Fusion Physics and Technology (Grant nr. 047.018.002). This work is part of the research programme `Fellowships for Young Energy Scientists' (YES!) of the Foundation for Fundamental Research on Matter (FOM), which is financially supported by the Netherlands Organisation for Scientific Research (NWO). The authors would like to thank E. Highcock, S. Moradi, G. Plunk, A. Schekochihin and E. Westerhof for stimulating discussions. Resources of the HELIOS supercomputer at IFERC-CSC are acknowledged. This research used computational resources at the National Research Scientific Computing Center, which is supported by the Office of Science of the U.S. Department of Energy under Contract No. DE-AC02-05CH11231. The authors are grateful to D. R. Mikkelsen for assistance. 

\footnotesize{
\bibliographystyle{unsrt}
\bibliography{jcitrin_PPCF_EPS_invited}}

\end{document}